\documentclass[twocolumn,trackchanges]{aastex701}

\usepackage{amsmath,amssymb}
\usepackage{graphicx}

\shorttitle{Topological Centrality in Gaia Open Clusters}
\shortauthors{Zaidyn et al.}

\begin{document}
\title{Topological Decomposition of Mass Segregation in Gaia Open Clusters}

\author[orcid=0009-0006-8505-7277]{Marat Zaidyn}
\affiliation{Department of Electronics and Astrophysics, Farabi University, 050040 Almaty, Kazakhstan}
\email{maratzayden@gmail.com}

\author[orcid=0009-0001-8784-4470]{Aknur Sakan*}
\affiliation{Department of Electronics and Astrophysics, Farabi University, 050040 Almaty, Kazakhstan}
\email[show]{aknursakan47@gmail.com}

\author[orcid=0000-0002-0045-5457]{Atilkhan Nodyarov}
\affiliation{Department of Electronics and Astrophysics, Farabi University, 050040 Almaty, Kazakhstan}
\email{nodyarov.atilkhan@gmail.com}

\author[orcid=0000-0002-9185-3185]{Aigerim Akniyazova}
\affiliation{Department of Electronics and Astrophysics, Farabi University, 050040 Almaty, Kazakhstan}
\email{aigerimakniyazova@gmail.com}

\author[orcid=0009-0002-5435-9740]{Alisher Zhunuskanov}
\affiliation{Department of Electronics and Astrophysics, Farabi University, 050040 Almaty, Kazakhstan}
\email{zhunuskanov_alisher@live.kaznu.kz}

\author[orcid=0000-0002-2512-3280]{Nurzhan Ussipov}
\affiliation{Department of Electronics and Astrophysics, Farabi University, 050040 Almaty, Kazakhstan}
\email{ussipov.nurzhan@kaznu.kz}

\begin{abstract}
Open clusters exhibit diverse spatial structures that provide insight into their formation and dynamical evolution. We investigate the relationship between stellar mass segregation, network topology, and dynamical age in 871 dynamically unevolved open clusters ($\tau = t/t_{\rm rh} \leq 1$) from Gaia-based catalogs. We construct three-dimensional k-nearest-neighbor graphs and measure Degree Centrality (DC), Closeness Centrality (CL), and Betweenness Centrality (BC). For each measure, we define a network-based mass segregation ratio (MSR) by comparing the mean centrality of the most massive stars with that of all cluster members. We test the stability of these measures across connectivity parameters $k=5$--20 and compare them with the classical 3D MSR, $\Lambda_{\rm MSR,3D}$, and dynamical age. All three network ratios show positive correlations with $\Lambda_{\rm MSR,3D}$, strongest for closeness centrality, while their correlations with dynamical age are weak. These results demonstrate that network analysis provides a complementary approach to studying stellar structure and mass segregation in open clusters.
\end{abstract}

\keywords{\uat{Open star clusters}{1160} --- \uat{Stellar dynamics}{1596} --- \uat{Gaia}{2360} --- \uat{Galaxy kinematics}{602}}

\section{Introduction}
\label{sec:intro}

Star clusters evolve through internal $N$-body interactions and external Galactic tides \citep{2003MNRAS.340..227B, 10.1111/j.1365-2966.2006.10122.x, Trenti_2010, 10.1111/j.1365-2966.2012.21668.x}. As stars exchange energy over time, the cluster naturally evolves toward energy equipartition \citep{1987degc.book.....S, 2010MNRAS.405..401D, 2011MNRAS.413.2345H, 10.1093/mnras/stt1521}. This continuous energy redistribution causes mass segregation \citep{1969ApJ...158L.139S, 2017MNRAS.464.1977W}. Massive stars sink toward the central core, while lower-mass members disperse outward. Although this spatial sorting is widely used as a standard diagnostic to assess the dynamical evolutionary stage of star clusters \citep{2009MNRAS.395.1449A, 2015ApJ...805...23C, Jadhav2021, Jiang_2024, 2026A&A...707A.311T, Zhao2026}, whether such signatures reflect purely dynamical evolution remains unresolved. In particular, the primordial spatial distribution of stars in young, dynamically unevolved clusters continues to be a subject of active debate \citep{Zhang2026againt, 2018MNRAS.473..849D, coenda2025global}.

In practice, mass segregation is widely quantified using minimum spanning tree (MST) metrics such as $\Lambda_{\rm MSR}$ \citep{2009MNRAS.395.1449A, 2018MNRAS.473..849D} and its sensitivity-enhanced refinement $\Gamma_{\rm MSR}$ \citep{olczak2011}.
However, interpreting these indicators is fraught with systematic biases: sample incompleteness and unresolved binaries can artificially mimic or erase segregation signals \citep{2009A&A...495..147A, 2011MNRAS.417.2300K}, while MSTs, despite being center-independent, trace global compactness rather than local crowdedness, rendering them blind to segregation distributed across multiple subclusters \citep{yu2017MSRsensetive}.
As a result, disparate spatial metrics often yield conflicting evolutionary signatures in substructured or non-spherical clusters \citep{2015MNRAS.449.3381P}, demonstrating that spatial segregation alone is an unreliable proxy for a cluster's dynamical history or degree of equipartition \citep{2017MNRAS.464.1977W}.

Because the MST is fundamentally a basic graph-theoretic construct, its limitations in disentangling multi-scale configurations naturally motivate a broader application of complex network science. In extragalactic astrophysics, network topologies are already used to map the cosmic web, where degree and betweenness centralities effectively trace dense environments and filamentary bridges \citep{hong2015network, Gausmann2024}. In star clusters and star-forming regions, this methodology is rapidly emerging: recent studies have used graph-based fractal dimensions to trace morphological evolution \citep{ussipov2024, xiaoying2025, akhmetali2026evolution}, mapped hierarchical cascades in molecular clouds \citep{Thomas2022}, and identified kinematic substructures via community detection \citep{Maxwell2025}.

In this paper, we construct adaptive $k$-nearest-neighbor ($k\text{NN}$) spatial graphs for each cluster \citep{inproceedings} and adapt network centrality metrics into structural mass segregation indicators \citet{hong2015network}.
We apply this topological framework to a sample of 871 open clusters compiled from \citet{2024yCat..36860042H} and \citet{2026AJ....172...85Z}. By comparing our network-based MSR against both standard 3D MST ratios ($\Lambda_{\rm MSR}$) and cluster dynamical ages ($\tau$), with a particular focus on dynamically unevolved systems ($\tau \le 1$), we test whether graph topology uncovers evolutionary and structural information inaccessible to traditional mass segregation diagnostics.

This paper is organized as follows: Section~\ref{sec:data} describes the cluster sample; Section~\ref{sec:method} details the $k\text{NN}$ network construction and centrality metrics; Section~\ref{sec:results} presents the statistical and residual analyses; and discusses the physical implications;  Section~\ref{sec:conclusions} summarizes our findings.

\section{Data}
\label{sec:data}

\begin{figure*}[ht!]
    \centering
    \includegraphics[width=\textwidth]{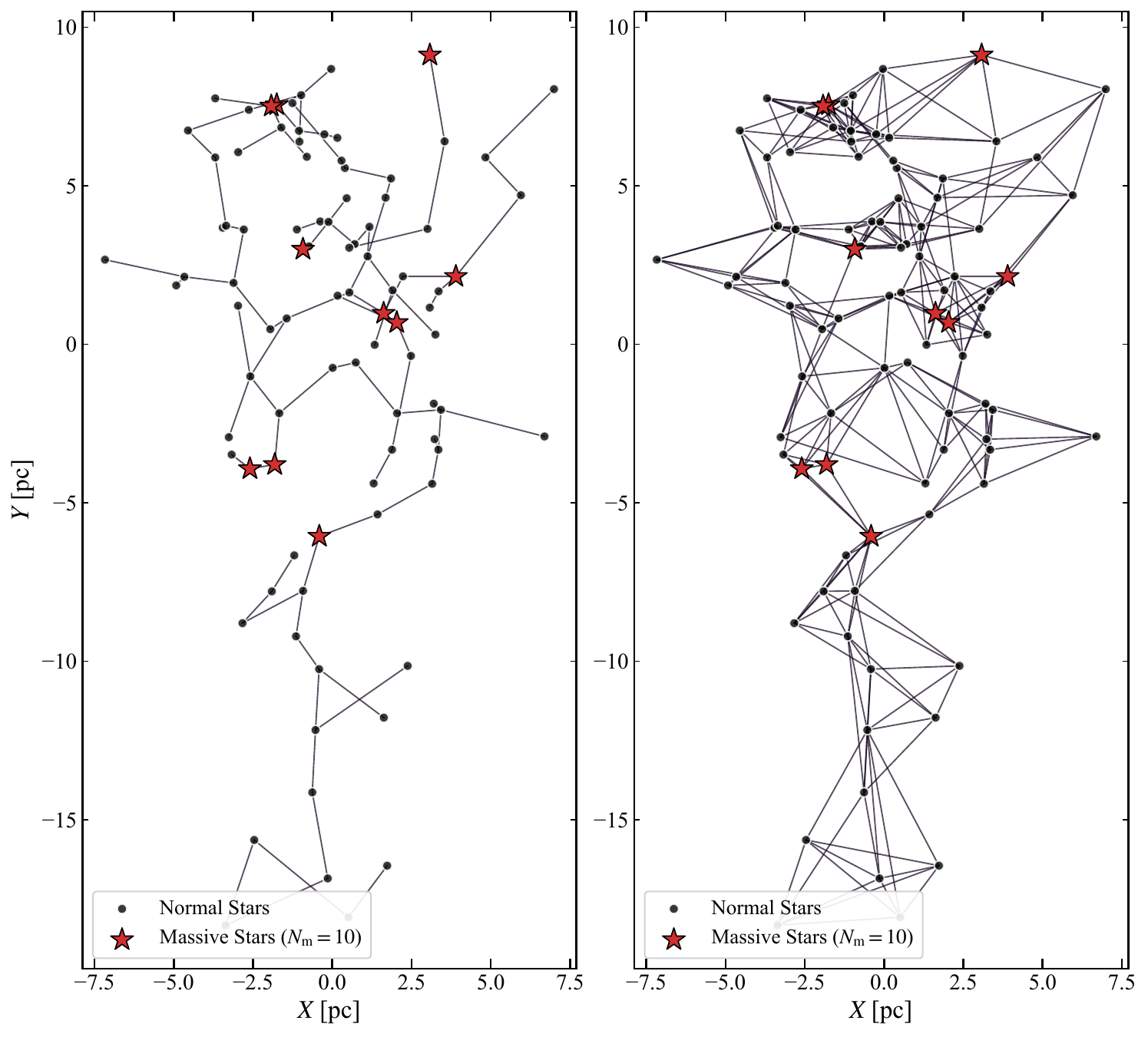}
    \caption{The $k$-NN ($k=5$, left panel) and MST for cluster (right panel) \texttt{OC\_0479} with median XY coordinates.}
    \label{fig1}
\end{figure*}

We base our analysis on the sample of 871 dynamically unevolved open clusters ($\tau \le 1$) cataloged by \citet{2026AJ....172...85Z}, which builds upon the astrometric and membership data of \citet{2024yCat..36860042H}. At the cluster level, we directly adopt the dynamical age ($\tau$) and the 3D MST-based MSR ($\Lambda_{\rm MSR,3D}$) computed by \citet{2026AJ....172...85Z}.

To construct the structural networks, we calculate the local 3D Cartesian coordinates for each  member star . Following the methodology of \citet{2026AJ....172...85Z}, we reconstruct stellar distances by applying a Bayesian correction to the Gaia parallaxes, weighting the individual measurement ($\varpi_{\rm ind}$) against the cluster mean ($\varpi_{\rm clus}$):
\begin{equation}
\varpi_{\rm corr} = \frac{\varpi_{\rm ind} \sigma_{\rm clus}^2 + \varpi_{\rm clus} \sigma_{\rm ind}^2}{\sigma_{\rm clus}^2 + \sigma_{\rm ind}^2}.
\end{equation}
Here, we set a floor for the cluster parallax uncertainty, $\sigma_{\rm clus} = \max(e_{\rm Plx}, 0.01)\;{\rm mas}$, to maintain numerical stability. The corrected distances ($d = 1000 / \varpi_{\rm corr}\;{\rm pc}$) and angular coordinates are then converted into a cluster-centric Cartesian frame $(X, Y, Z)$ to compute the pairwise Euclidean distances for the network edges. Because our membership cuts and parallax corrections align precisely with \citet{2026AJ....172...85Z}, we refer the reader to that work for full details on the sample selection.

\section{Methodology}
\label{sec:method}

\begin{figure*}[ht!]
    \centering
    \includegraphics[width=\textwidth]{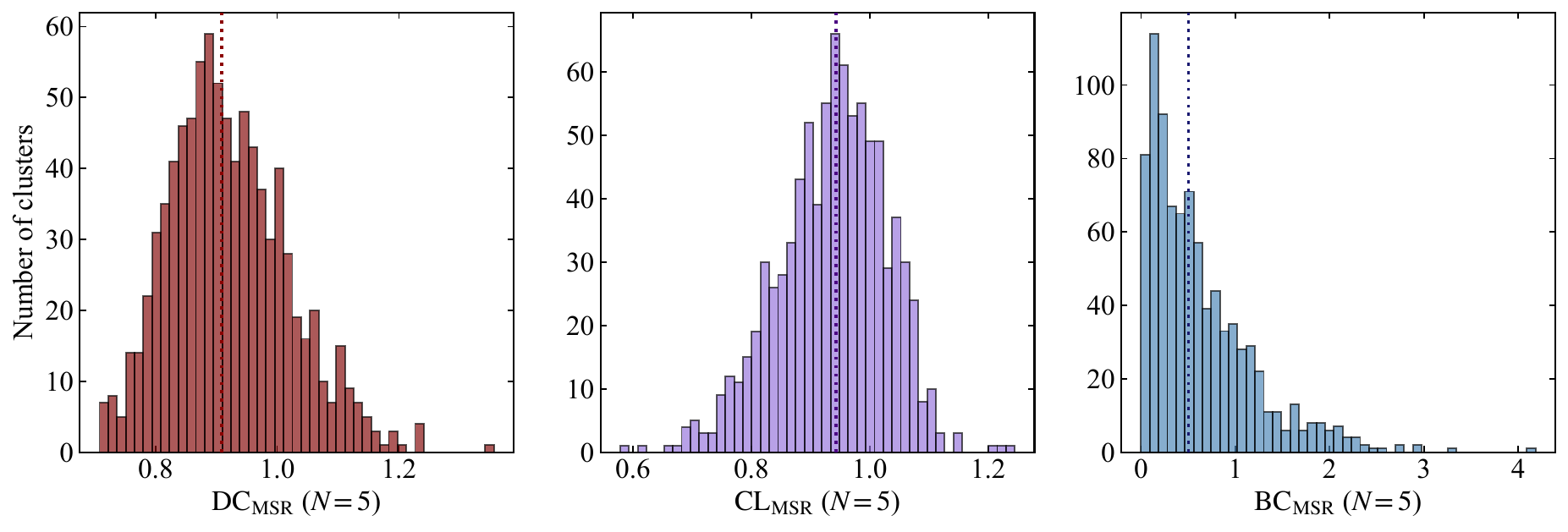}
    \caption{Distributions of the degree ($\mathrm{DC}_{\mathrm{MSR}}$), closeness ($\mathrm{CL}_{\mathrm{MSR}}$), and betweenness ($\mathrm{BC}_{\mathrm{MSR}}$) MSR for $N=5$ most massive stars across the 871 open clusters. The vertical dotted lines indicate the median values of each distribution: 0.91, 0.94 and 0.50 respectively.}
    \label{fig2}
\end{figure*}

\begin{figure*}[ht!]
    \centering
    \includegraphics[width=\textwidth]{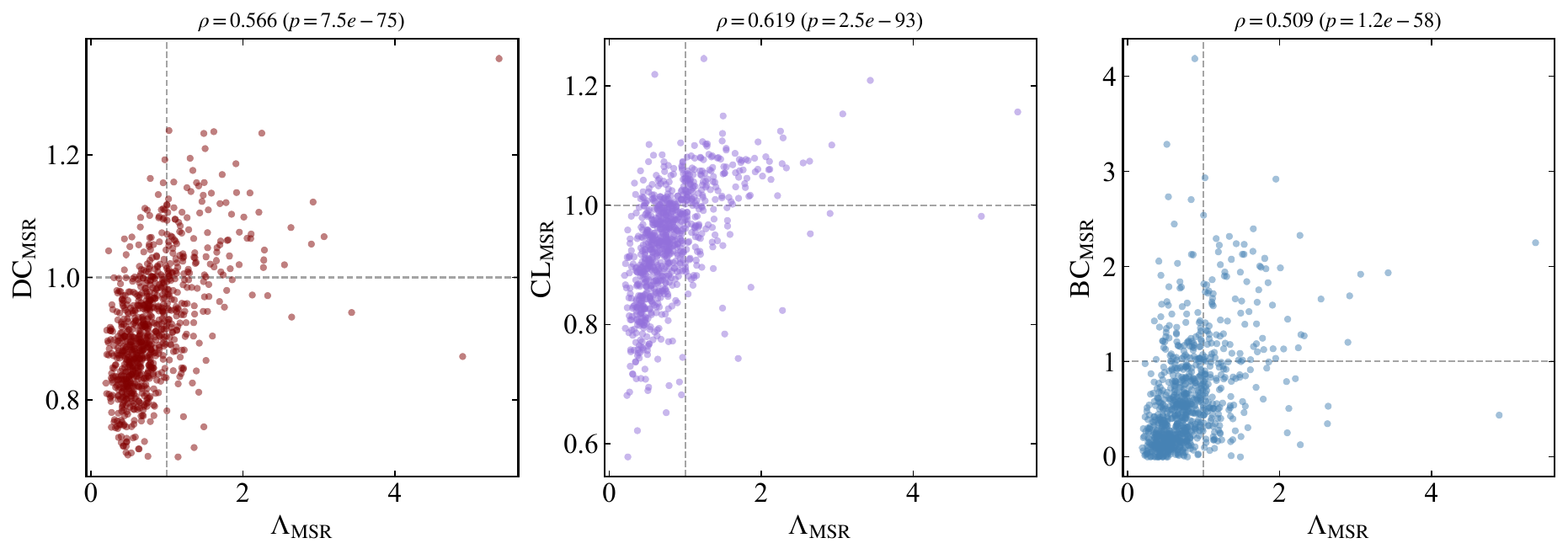}
    \caption{Comparison of network-based MSR derived from degree ($\mathrm{DC}_{\mathrm{MSR}}$), closeness ($\mathrm{CL}_{\mathrm{MSR}}$), and betweenness centrality ($\mathrm{BC}_{\mathrm{MSR}}$) with the conventional three-dimensional spatial mass segregation parameter, $\Lambda_{\mathrm{MSR,3D}}$. The dashed horizontal line marks $\mathrm{MSR}=1$, while the dashed vertical line indicates $\Lambda_{\mathrm{MSR,3D}}=1$, separating clusters with and without evidence of spatial mass segregation. Spearman rank correlation coefficients are shown in each panel.}
    \label{fig4}
\end{figure*}

To robustly capture the spatial topology of open clusters, we construct an undirected $k$NN graph, $G=(V,E)$, for each cluster. Based on the 3D Cartesian coordinates, we first compute the directed nearest-neighbor distances. Let $K$ be the asymmetric distance matrix where $K_{ij} = d_{ij}$ if star $j$ is among the $k$ nearest neighbors of star $i$, and $K_{ij} = 0$ otherwise ($d_{ij}$ is the Euclidean distance). 

To ensure the reciprocity of the structural topology, we symmetrize the network into an undirected graph. The final weighted adjacency matrix $W$ of graph $G$ is defined by taking the union of mutual neighbors:

\begin{equation}
W_{ij} = 
\begin{cases} 
d_{ij}, & \text{if } K_{ij} > 0 \text{ or } K_{ji} > 0 \\
0, & \text{otherwise}
\end{cases}
\end{equation}
The unweighted graph topology, which describes the pure connection paths, is then strictly given by the binary adjacency matrix $A$, where $A_{ij} = 1$ if $W_{ij} > 0$, and $A_{ij} = 0$ otherwise. Figure~\ref{fig1} illustrates the topological differences between the MST and k-NN graphs in the X--Y projection.

We evaluate the topological importance of each member star using three standard network centralities, computed via the \texttt{networkx} library. Degree Centrality (DC) measures local stellar density based on direct connections:
\begin{equation}
{\rm DC}_i = \frac{1}{n-1} \sum_{j \neq i} A_{ij},
\end{equation}
where $n$ is the total number of vertices. Closeness Centrality (CL) measures the topological proximity to the entire cluster:
\begin{equation}
{\rm CL}_i = \frac{n-1}{\sum_{j \neq i} h_{ij}},
\end{equation}
where $h_{ij}$ is the unweighted shortest-path distance (number of hops) between $i$ and $j$ on the binary matrix $A$. Betweenness Centrality (BC) identifies structural bottlenecks:
\begin{equation}
{\rm BC}_i = \frac{1}{(n-1)(n-2)} \sum_{s \neq t \neq i} \frac{g_{st}(i)}{g_{st}},
\end{equation}
where $g_{st}$ is the total number of shortest paths between nodes $s$ and $t$, and $g_{st}(i)$ is the number of those paths passing through $i$. The prefactor normalizes the metric by the number of possible source-target pairs in an undirected graph. BC is evaluated on the weighted matrix $W$, ensuring that the shortest paths trace the true metric distance through the cluster.

To quantify structural mass segregation, we compute the ratio of the mean centrality of the $N_{\rm massive}$ most massive stars to the mean centrality of the entire cluster. Let $C \in \{{\rm DC, CL, BC}\}$ denote a given centrality measure. The network-based MSR is defined as:

\begin{equation}
\Lambda_{C} = \frac{\langle C \rangle_{\rm massive}}{\langle C \rangle_{\rm all}},
\end{equation}
which we calculate for $N_{\rm massive} \in \{5, 10, 20\}$. A value of $\Lambda_{C} > 1$ indicates classical mass segregation, meaning the massive stars reside in topologically central, highly connected regions. Conversely, $\Lambda_{C} < 1$ indicates inverse segregation, where massive stars occupy the network periphery. Crucially, betweenness peaks at intermediate bottlenecks rather than the cluster center. Core concentrated massive stars thus yield low betweenness, making $\Lambda_{\rm BC} < 1$ a signature of core localization rather than inverse segregation. We therefore treat $\Lambda_{\rm DC}$ and $\Lambda_{\rm CL}$ as standard segregation metrics, while reserving $\Lambda_{\rm BC}$ specifically to probe bottleneck occupancy.

\begin{figure*}[ht!]
    \centering
    \includegraphics[width=\textwidth]{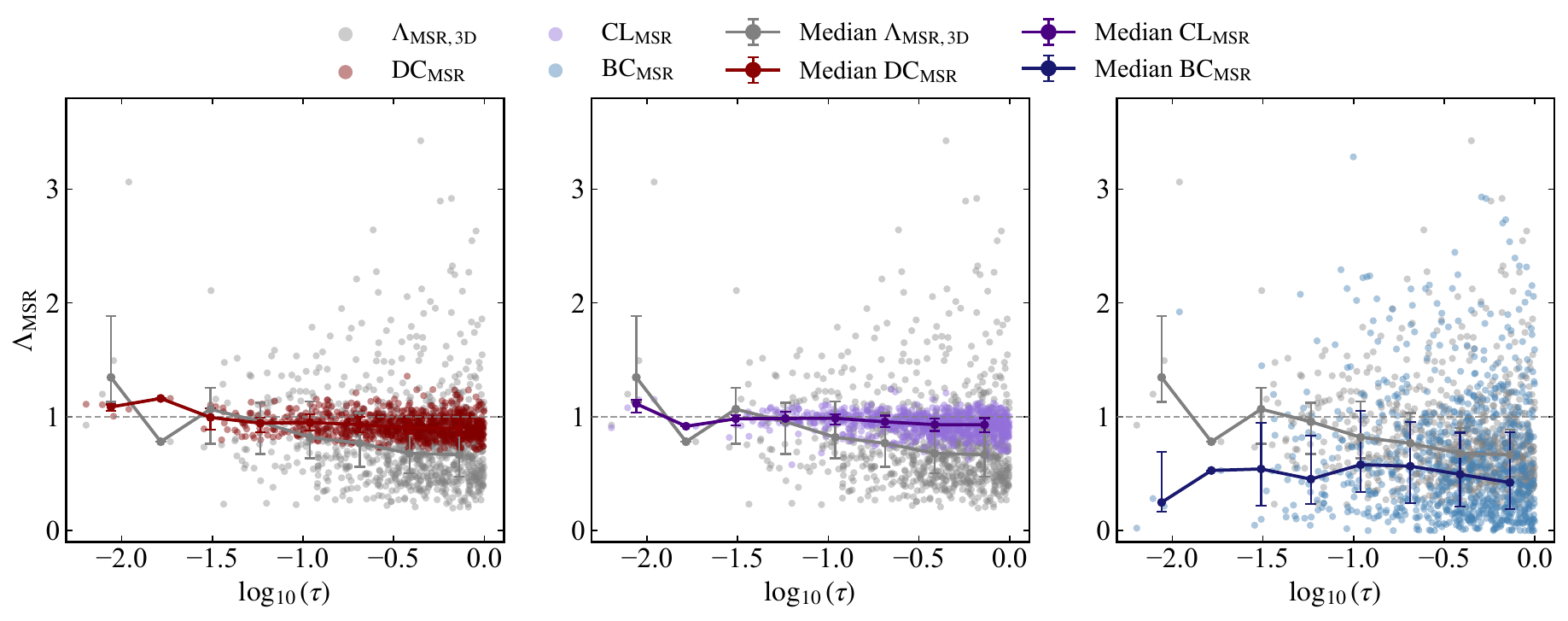}
    \caption{ Dependence of the degree ($\mathrm{DC}_{\mathrm{MSR}}$), closeness ($\mathrm{CL}_{\mathrm{MSR}}$), and betweenness ($\mathrm{BC}_{\mathrm{MSR}}$) on the logarithm of the cluster dynamical age, $\log_{10}(\tau$). Spearman rank correlation coefficients are shown in the title, illustrating the strength and direction of the association between network-based mass segregation and dynamical age.}
    \label{fig5}
\end{figure*}

\section{Results and Discussion } \label{sec:results}

To establish a baseline, we measure the network MSR for the $N_{\rm massive} = 5$ most massive stars across all 871 clusters (Figure~\ref{fig2}). Degree (${\rm DC}_{\rm MSR}$) and closeness (${\rm CL}_{\rm MSR}$) centralities yield symmetric, quasi-Gaussian distributions tightly centered marginally below unity ($\tilde{x}_{\rm DC} = 0.91$, $\tilde{x}_{\rm CL} = 0.94$), consistent with the prevalence of inverse segregation reported by \citet{2026AJ....172...85Z}. We adopt k=10 as the baseline connectivity parameter, its robustness and selection are discussed in Appendix~\ref{sec:appendix}.

In contrast, betweenness centrality (${\rm BC}_{\rm MSR}$) is strongly right-skewed ($\tilde{x}_{\rm BC} = 0.56$). As established in Section~\ref{sec:method}, $\Lambda_{\rm BC} < 1$ does not indicate inverse segregation: path redundancy suppresses betweenness at density peaks, while bottlenecks concentrate in intermediate shells. Consequently, core-concentrated massive stars inherently yield low betweenness by construction.

We compare the network ratios to the 3D spatial segregation parameter $\Lambda_{\rm MSR,3D}$ (Figure~\ref{fig4}). All three metrics correlate positively with spatial segregation along a strict hierarchy: closeness is strongest ($\rho = 0.619$), degree intermediate ($\rho = 0.566$), and betweenness weakest ($\rho = 0.509$).

This monotonic decline reflects distinct topological dimensions rather than metric redundancy. Closeness couples local connectivity to global shortest paths, overlapping most directly with the geometric compactness traced by the MST. Degree probes purely local density, with its dynamic range compressed by the adaptive $k$NN construction. Betweenness measures bottleneck occupancy a property orthogonal to compactness, as path redundancy suppresses central values. This hierarchy confirms that CL, DC, and BC extract complementary structural information, consistent with the multiscale frameworks of \citet{yu2017MSRsensetive} and \citet{hong2015network}.

Finally, Figure~\ref{fig5} traces the network ratios against dynamical age ($\log_{10}\tau$). While ${\rm DC}_{\rm MSR}$ and ${\rm CL}_{\rm MSR}$ mirror the broad secular decline of $\Lambda_{\rm MSR}$ \citep{2026AJ....172...85Z}, their decay is shallower and far more stable. Near $\log_{10}\tau \sim 1.5$--$1.7$, where \citet{2026AJ....172...85Z} report an abrupt drop and rebound, the network metrics undergo genuine peaks and subsequent rebounds, but with heavily attenuated amplitudes. Concurrently, ${\rm BC}_{\rm MSR}$ persists strictly below unity across all ages, tracing a subdued, inverted counterpart to $\Lambda_{\rm MSR}$.

The smoother evolution of DC and CL is consistent with the interpretation above: these two centralities average over local and global connectivity, suppressing the sensitivity of MST to the specific configuration of the ten most massive stars. The mirrored behavior of BC follows from the same topological selection discussed in Section~\ref{sec:method}: as a cluster relaxes and massive stars sink toward the core, BC decreases by construction, so $\Lambda_{\rm BC}$ tracks $\Lambda_{\rm MSR}$ with opposite sign.

\section{Conclusions}
\label{sec:conclusions}

Applying network centrality metrics to 871 dynamically unevolved Gaia open clusters ($\tau \leq 1$) demonstrates that network topology effectively traces mass segregation. Centrality ratios correlate positively with the standard 3D MST metric $\Lambda_{\mathrm{MSR,3D}}$ along a monotonic hierarchy closeness ($\rho = 0.619$), degree ($\rho = 0.566$), and betweenness ($\rho = 0.509$) which remains stable across $k = 5$-$20$ and confirms a structural decomposition spanning global accessibility, local crowding, and bottleneck occupancy. Ultimately, network centrality enriches classical spatial methods by resolving single-metric geometric compactness into complementary topological dimensions: closeness offers the closest proxy to $\Lambda_{\mathrm{MSR,3D}}$, while betweenness reveals orthogonal bottleneck structures inaccessible to the MST.

\begin{acknowledgments}
This work has made use of data from the European Space Agency (ESA) mission 
\textit{Gaia} (\url{https://www.cosmos.esa.int/gaia}), processed by the 
\textit{Gaia} Data Processing and Analysis Consortium (DPAC). This research 
was funded by the Science Committee of the Ministry of Science and Higher 
Education of the Republic of Kazakhstan through grant AP32315949.
\software{Astropy \citep{Astropy2022}, NumPy \citep{2020Natur.585..357H},
SciPy \citep{2020NaMet..17..261V}, Matplotlib \citep{2007CSE.....9...90H}, Networkx \citep{hagberg2008exploring}.}
\end{acknowledgments}

\section*{AI Disclosure}
During the preparation of this work, the authors utilized large language models (LLMs) 
to assist with English language polishing and proofreading. The authors reviewed, 
edited, and take full responsibility for the scientific content and integrity of the 
final manuscript.

\bibliography{references}
\bibliographystyle{aasjournalv7}

\begin{appendix}
\section{Robustness of the Mass Segregation Ratio}
\label{sec:appendix}

\begin{figure*}[ht!]
    \centering
    \includegraphics[width=\textwidth]{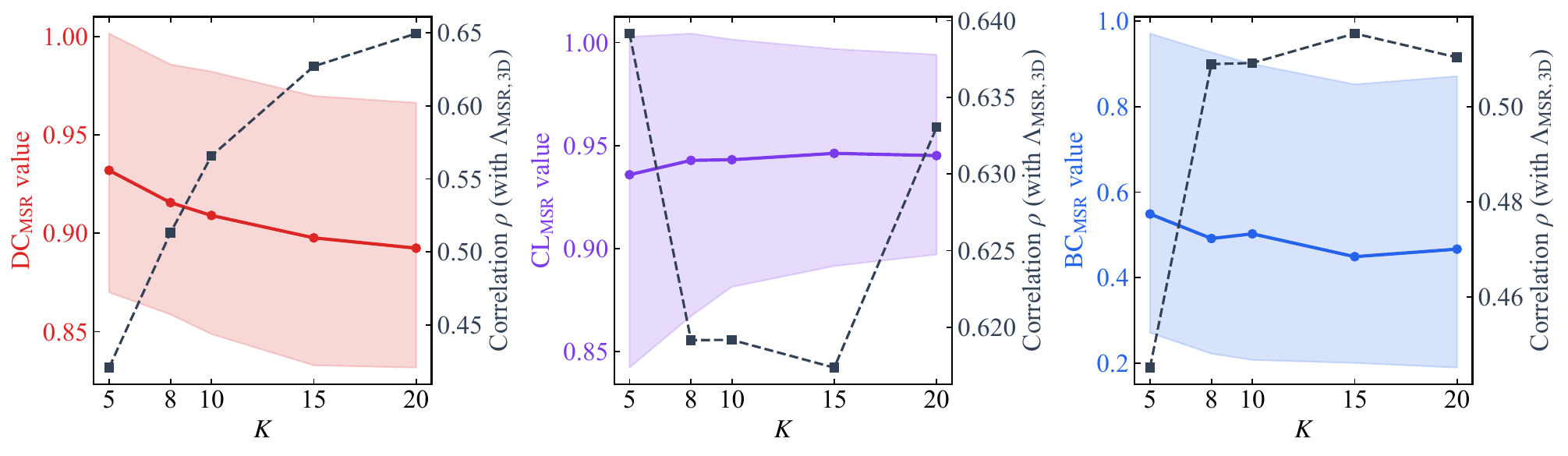}
    \caption{To test whether the graph connectivity parameter affects our conclusions, we vary $k \in \{5, 8, 10, 15, 20\}$ for $N=5$ most massive stars and evaluate both the cluster MSR distributions (with shaded regions denoting the interquartile ranges) and their Spearman rank correlation with the classical $\Lambda_{\rm MSR, 3D}$.}
    \label{fig3}
\end{figure*}

To ensure these structural measurements are stable, we tested our method by varying the network neighbor parameter, $k$, from 5 to 20, as illustrated in Figure~\ref{fig3}. Closeness centrality remains highly stable across all tested $k$ values. Degree and betweenness drop slightly at higher $k$, but their correlation with spatial segregation actually improves. This confirms our proposed method is mathematically robust and not simply an artifact of choosing a specific graph parameter. Because the structural interpretation remains consistent regardless of the exact parameter, we adopt $k=10$ as the standard value for all subsequent analyses.   

\end{appendix}
\end{document}